\documentclass[prl,letterpaper,english,reprint,nofootinbib,aps,superscriptaddress,showpacs,showkeys]{revtex4-1}

\usepackage{babel,calc,amsmath,amsthm,amssymb,graphicx,subfigure,xcolor,comment}
\usepackage{mathdots}
\usepackage[T1]{fontenc}
\usepackage[unicode=true]{hyperref}
\usepackage{booktabs}
\usepackage{threeparttable}

\hypersetup{
     colorlinks=true,       		
     linkcolor=blue,          	
     citecolor=red,            
     urlcolor=magenta,           	
 }

\newtheorem*{theorem*}{Theorem}

\newtheorem*{corollary*}{Corollary}

\newtheorem*{lemma*}{Lemma}

\newtheorem*{proposition*}{Proposition}
\theoremstyle{definition}

\newtheorem*{definition*}{Definition}
\theoremstyle{remark}

\newtheorem*{remark*}{Remark}

\newcommand{\tr}{\rm tr}

\begin{document}

\title{ Mapping Criteria of Nonlocality-Steerability in Qudit-Qubit Systems and Steerability-Entanglement in Qubit-Qudit Systems }

\author{Changbo Chen}\thanks{These authors contributed equally.}
 \affiliation{Chongqing Key Laboratory of Automated Reasoning and Cognition, Chongqing Institute of Green and Intelligent Technology, Chinese Academy of Sciences, Chongqing 400714, People's Republic of China}\email{chenchangbo@cigit.ac.cn}

\author{Changliang Ren}\thanks{These authors contributed equally.}
\affiliation{Center for Nanofabrication and System Integration, Chongqing Institute of Green and Intelligent Technology, Chinese Academy of Sciences, Chongqing 400714, People's Republic of China}\email{ renchangliang@cigit.ac.cn}\affiliation{CAS Key Laboratory of Quantum Information, University of Science and Technology of China, Hefei 230026, PR China}

\author{Xiang-Jun Ye}
\affiliation{CAS Key Laboratory of Quantum Information, University of Science and Technology of China, Hefei 230026, People's Republic of China}

\author{Jing-Ling Chen}
 \affiliation{Theoretical Physics Division, Chern Institute of Mathematics, Nankai University,
 Tianjin 300071, People's Republic of China}\email{chenjl@nankai.edu.cn}
 \affiliation{Centre for Quantum Technologies, National University of Singapore,
 3 Science Drive 2, Singapore 117543}
%
%

\date{\today}
\begin{abstract}
Entanglement, quantum steering and nonlocality are distinct quantum correlations which are the resources behind various of quantum information and quantum computation applications. However, a central question of determining the precise quantitative relation among them is still unresolved. Here, we present a  mapping criterion between Bell nonlocality and quantum steering in bipartite qudit-qubit system,
as well as a  mapping criterion between quantum steering and quantum entanglement in bipartite qubit-qudit system, starting
from the fundamental concepts of quantum correlations.
Precise quantitative mapping criteria are derived analytically.
Such mapping criteria are independent of the form of the state.
In particular, they cover several previous well-known research results which are only special cases in our simple mapping criteria.

\end{abstract}


\maketitle


The distinctive non-classical features of quantum physics were
first discussed in the seminal paper \cite{Einstein} by A. Einstein, B. Podolsky
and N. Rosen (EPR) in 1935, which indicated that there were some conflicts between quantum
mechanics and local realism. Immediately, The EPR paper provoked an interesting response from
Schr\"{o}dinger \cite{Schrodinger1,Schrodinger}, who introduced the notion of entanglement and steering. Three fundamental definition, ``quantum entanglement" \cite{Horodecki}, ``EPR steering" \cite{Cavalcanti}, and ``Bell nonlocality" \cite{Brunner}  were intuitively elaborated, which have since opened an epoch of unrelenting exploration of quantum correlations. Entanglement and Bell non-locality have attained flourishing developments since 1964, while EPR steering had even lacked a rigorous formulation  until the work in 2007 due to Wiseman \emph{et al.} \cite{Wiseman}. Over 80 years investigation, physicists have scrupulously distinguished the notions and clarified the concepts out of chaos.
These concepts have nowadays become the center of quantum foundations and have found themselves many practical applications
in modern quantum information theory ranging from quantum key distribution \cite{Ekert,Gisin,Acin,Branciard,Kocsis},
communication complexity \cite{Brukner,Piani}, cloning of correlations \cite{Piani1,Luo}, quantum metrology \cite{Modi},
quantum state merging \cite{Oppenheim,Cavalcanti1}, remote state preparation \cite{Dakic}, and random number generation \cite{Pironio}.

Through decades of investigation, a great number of fruitful results on characterizing the properties of
these quantum correlations have been obtained \cite{Schrodinger,Bell,Werner,Greenberger,Jones}.
According to the hierarchy of non-locality, the set of EPR steerable states is a strict subset of entangled states
and a strict superset of Bell nonlocal states \cite{Jones}.
In simplicity, the strongest concept is Bell non-locality,
which implies non-classical correlations that cannot be described by local hidden variable theory (LHV);
quantum steering describes correlations beyond ones constrained by local hidden state theory (LHS);
the strictly weaker concept is that of nonseparability
or entanglement, where a nonseparable state is one that its joint probability cannot be simulated by any separable model (SPM).
However, the above is basically the whole knowledge of the relations among these three different quantum correlations.
Especially, there are very few quantitative results on the relation of such quantum correlations.
Quantitatively determining their difference and relation is an important task:
it helps deeply understanding the nonclassical physics described by quantum mechanics
and provides a verification of them in terms of their usefulness for various quantum information applications.
In this paper, a mapping criterion of entanglement-steerability in qudit-qubit system and
a mapping criterion of steerability-nonlocality in qubit-qudit system
were derived from their fundamental definitions respectively.
As a result,
we are able to prove that a difficultly-verified quantum correlation can be translated into an easily-verified problem.
This result connects the previous research of detecting Bell's nonlocality by quantum steering inequality in \cite{Chen,Cavalcanti2} to the relatively new research direction of steering.
It is shown that part of these previous known result in \cite{Chen,Cavalcanti2} is only a special case for 2-qubit system in our simple mapping criterion.
Moreover, the perspective in this research supply a novel simple way of exploring the relation of such quantum correlations quantitatively.


\emph{Preliminary notions.-}Consider a bipartite scenario composed by Alice and Bob sharing an arbitrary quantum state $\tau_{AB}$. Suppose Alice performs measurement ${A}$ with outcome $a$ and Bob performs measurement $B$ with outcome $b$, then these outcomes are thus in general governed by a joint probability distribution $P(a,b\mid A,B,\tau_{AB})$, where this joint probability distribution predicted by quantum theory is defined by:
\begin{eqnarray}\label{QM}
P(a,b\mid A,B,\tau_{AB})=\mathrm{Tr}[({\Pi}^{A}_{a}\otimes {\Pi}^{B}_{b})\tau_{AB}],
\end{eqnarray}
where ${\Pi}^{A}_{a}$
and ${\Pi}^{B}_{b}$ are the projective operators for Alice and Bob respectively.

\textbf{ Definition 1. } If the joint probability
  satisfies
\begin{eqnarray}\label{LHV}
  P(a, b|A, B, \tau_{AB})&=&\int P(a|A,\xi)P(b|B,\xi)P_{\xi} d\xi,
\end{eqnarray}
for any measurements $A$ and $B$,
$\tau_{AB}$ has a local hidden variable (LHV) model.

\textbf{ Definition 2. }If the joint probability and the marginal probability satisfy
\begin{eqnarray}\label{LHS}
  P(a, b|A, B, \tau_{AB})&=&\int P(a|A,\xi)P_{Q}(b|B,\xi)P_{\xi} d\xi,\\
P_{Q}(b|B,\xi)&=&\mathrm{Tr}[{\Pi}^{B}_{b}\rho^{B}_{\xi}],
\end{eqnarray}
for any measurements $A$ and $B$,
$\tau_{AB}$ has a local hidden state (LHS) model, where $\int P_{\xi}\rho^{B}_{\xi}\mathrm{d}\xi=\mathrm{Tr}_{A}[\tau_{AB}]$.

\textbf{ Definition 3. }If the joint probability and the marginal probability satisfy
\begin{eqnarray}\label{SP}
  	P(a, b|A, B, \tau_{AB})&=&\int P_{Q}(a|A,\xi)P_{Q}(b|B,\xi)P_{\xi} d\xi,\\
	P_{Q}(a|A,\xi)&=&\mathrm{Tr}[{\Pi}^{A}_{a}\rho^A_{\xi}],\\
	P_{Q}(b|B,\xi)&=&\mathrm{Tr}[{\Pi}^{B}_{b}\rho^B_{\xi}],
\end{eqnarray}
for any measurements $A$ and $B$,
$\tau_{AB}$ has a separable model (SPM).

\begin{figure}
\includegraphics[width=0.3\textwidth]{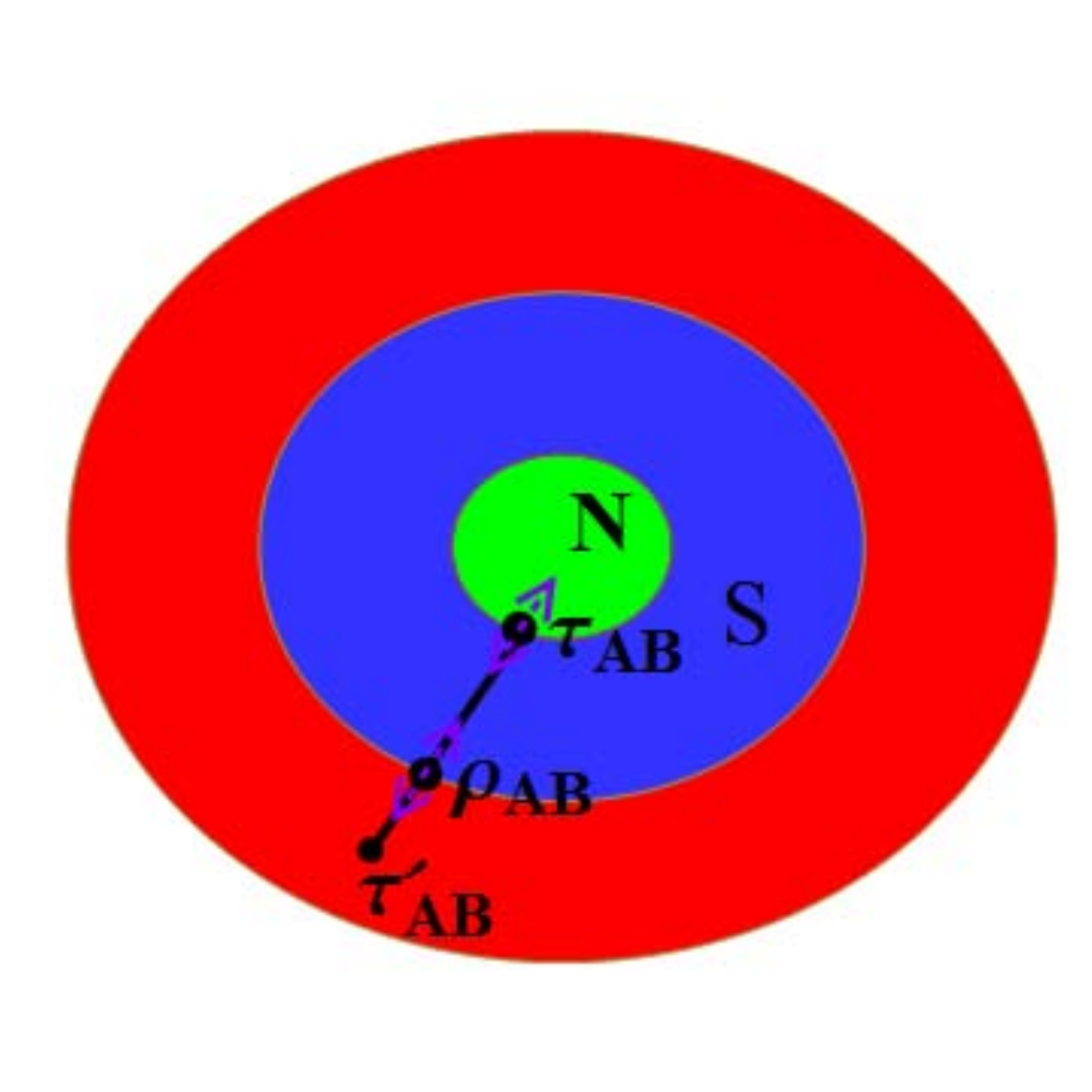}
 \caption{\label{fig1}(Color online) The Venn diagram of a mapping relation between quantum nonlocality and quantum steering in Theorem 1. The states in Green region is nonlocal. And those states in the Blue region is steerable but not nonlocal. All of the states which can be described by LHS model are in the Red region. The state $\rho_{AB}$ is mixed by an arbitrary unsteerable state $\tau'_{AB}$ and the other arbitrary state $\tau_{AB}$. Theorem 1 gives a mapping criterion between $\rho_{AB}$ and $\tau_{AB}$. The purple arrows show that, if $\rho_{AB}$ is EPR steerable, then $\tau_{AB} $ is Bell nonlocal, equivalently if $\tau_{AB} $ is not nonlocal, then $\rho_{AB}$ is unsteerable. }
\end{figure}

\emph{Mapping criterion between Bell nonlocality and quantum steering.-} In what follows we present a mapping criterion between Bell nonlocality and quantum steering. A curious quantum phenomenon directly connecting these two different types of quantum correlations was proposed. We find that Bell nonlocal states can be constructed from some EPR steerable states, which indicates that Bell's nonlocality can be detected indirectly through EPR steering
(see Fig. (1)), and offers a distinctive way to study Bell's nonlocality. The result can be expressed as the following
theorem.

\textbf{ Theorem 1. }In a bipartite qudit-qubit system, we define a map ${\cal M}: \tau_{AB}\rightarrow \mu \; \tau_{AB} +(1-\mu) \tau'_{AB}$, $0\leq \mu\leq 1,$ where
  $\tau_{AB}$ is an arbitrary bipartite
qudit-qubit state shared by Alice and Bob,
while $\tau_{AB}'$ is a  bipartite
qudit-qubit state constructed in such a way that whenever $\tau_{AB}$ has a LHV model:
\begin{eqnarray}\label{LHV}
	P(a,b|A, B, \tau_{AB} )=\int P(a|A,\xi )P(b|B,\xi )P_{\xi } d\xi,
\end{eqnarray}
$\tau_{AB}'$ also has a LHV model:
\begin{eqnarray}\label{LHVb}
	P(a,b|A, B, \tau_{AB}' )=\int P'(a|A,\xi )P'(b|B,\xi )P_{\xi }d\xi.
\end{eqnarray}
Note that Eqs.~(\ref{LHV}) and~(\ref{LHVb}) contain the same $P_{\xi}$.
If there exists a range of $\mu$ such that $r_x^2+r_y^2+r_z^2\leq 1$ holds
for any probability distributions $0\leq P(a|A,\xi)\leq 1$, $0\leq P(b|B,\xi)\leq 1$,
where
$A$ is an arbitrary projective measurement, $B\in \{x,y,z\}$, and
\begin{equation}
  \label{eqs:rxyz}
\begin{array}{rcl}
 r_{x}&=&\frac{2\eta(x)}{\wp(a|A,\xi)}-1,\\
 r_{y}&=&\frac{2\eta(y)}{\wp(a|A,\xi))}-1, \\
 r_{z}&=&\frac{2\eta(z)}{\wp(a|A,\xi)}-1,
 \end{array}
\end{equation}
$\eta(B)=(\mu P(a|A,\xi)P(0|B,\xi)+(1-\mu)P'(a|A, \xi)P'(0|B, \xi))~\mbox{with}~B\in \{x,y,z\}$,
and
$\wp(a|A,\xi)=\mu P(a|A,\xi )+(1-\mu)P'(a|A,\xi)$,
then when $\mu$ falls into this range,
one can construct a LHS model for $\rho_{AB}={\cal M}(\tau_{AB})$.

Proof. Let the measurement settings at Bob's side be picked out as ${x, y,z}$.
Since the state $\tau_{AB}$ has a LHV model description, based on Eq.~(\ref{LHV})
we explicitly have (with $B= x, y, z$)
	\begin{eqnarray}\label{ELHV}
		P(a,0|A, B, \tau_{AB} )&=&\int P(a|A,\xi)P(0|B,\xi )P^{}_{\xi }d\xi,\nonumber\\
		P(a,1|A, B, \tau_{AB} )&=&\int P(a|A,\xi)P(1|B,\xi )P^{}_{\xi }d\xi.
	\end{eqnarray}

We now turn to study the EPR steerability of $\rho_{AB}$. After
Alice performs the projective measurement on her qubit, the state
$\rho_{AB}$ collapses to Bob's conditional states (unnormalized) as
\begin{eqnarray}\label{pp}
        \tilde{\rho}^{A}_a=\tr_A[({{\Pi}} ^{A}_{a}
        \otimes \openone) \rho_{AB}], \;\;\; a=0, ..., d-1.
\end{eqnarray}
To prove that there exists a LHS model for $\rho_{AB}$,
it suffices
to prove that, for any projective measurement ${{\Pi}} ^{A}_a$
and outcome $a$, one can always find a hidden state ensemble
$\{\wp_{\xi} \rho_{\xi} \}$ and the conditional probabilities
$\wp(a|A,\xi)$, such that the relation
\begin{eqnarray}\label{LHS1}
        &&\tilde{\rho}^{A}_a=\int \wp(a|A,\xi)\rho_{\xi}
        \wp_{\xi} d\xi,
\end{eqnarray}
is always satisfied. Here $\xi$ is a local hidden variable,
$\rho_{\xi}$ is a hidden state, $\wp_{\xi}$ is a probability density function
and
$\wp(a|A,\xi)$ are probabilities satisfying
$\int \wp_{\xi}d\xi=1$ and $\sum_a \wp(a|A,\xi) =1$.
Indeed, if Eq. (\ref{LHS1}) is satisfied,
then Eq. (\ref{LHS}) holds by calculating $\tr[{\Pi}^{B}_{b}\tilde{\rho}^{A}_a]$.


Each $\rho_{\xi}$ is a $2\times 2$ density matrix which can be written in the form of $\frac{\openone+\vec{\sigma}\vec{r}_{\xi}}{2}$,
  where $\openone$ is the $2\times 2$ identity matrix, $\vec\sigma=(\sigma_x,  \sigma_y, \sigma_z)$ is
  the vector of the Pauli matrices, and $\vec{r}_{\xi}=(r_x,r_y, r_z)$ is
  the Bloch vector satisfying $r_x^2+r_y^2+r_z^2\leq 1$.
  A  solution of Eq. (\ref{LHS1}) can be given as follows:
\begin{eqnarray}\label{LHS2}
        &&\wp(a|A,\xi)=\mu P(a|A,\xi )+(1-\mu)P'(a|A,\xi) \;\;\nonumber \\&&\wp_{\xi}={P}_{\xi },
         \rho_\xi=\frac{\openone+\vec{\sigma}\cdot \vec{{r}}_{\xi }}{2},
\end{eqnarray}
where the hidden state $\rho_\xi$ has been parameterized in the
Bloch-vector form, with $\vec{r}_{\xi}=(r_x,r_y, r_z)$
defined in Eq.~(\ref{eqs:rxyz}).
The assumption that $|\vec{r}_{\xi }|\leq 1$ ensures that $\rho_\xi$ is a density matrix.

By substituting Eq. (\ref{LHS2}) into Eq. (\ref{LHS1}), we obtain
	\begin{eqnarray}\label{LHS3}
		\tilde{\rho}^{A}_a=\int \wp(a|A,\xi)
		\frac{\openone+\vec{\sigma}\cdot \vec{{r}}_{\xi }}{2}{P}_{\xi }d\xi.
	\end{eqnarray}
        To prove the theorem is to verify that the relation (\ref{LHS3}) is satisfied.

	Let us calculate the left-hand side of Eq. (\ref{LHS3}). One has
 \begin{eqnarray*}\label{left1}
   \tilde{\rho}^{A}_a
   &=&\tr_A[({\Pi} ^{A}_{a} \otimes \openone) \rho_{AB}] \nonumber\\
					 &=&\mu\;\tr_A[({\Pi} ^{A}_{a} \otimes \openone)\tau_{AB} ]
					   +(1-\mu)\tr_A[({\Pi} ^{A}_{a} \otimes \openone)\tau_{AB}' ].
				 \end{eqnarray*}
 For convenience, let us denote the $2\times
				 2$ matrix $\tilde{\rho}^{A}_a$ as
				 \begin{eqnarray*}\label{left}
					 \tilde{\rho}^{A}_a=\begin{bmatrix}
						  {\nu}_{11}&{\nu}_{12} \\
						   {\nu}_{21}&{\nu}_{22}
					 \end{bmatrix},
				 \end{eqnarray*}
				 and calculate its each element. Obviously, we get
					 \begin{eqnarray*}\label{element11}
						 {\nu}_{11}&=&\tr[{\Pi}^{z}_{0}\; \tilde{\rho}^{A}_a]\nonumber\\
						 &=&\mu\; P(a,0|A,z,\tau_{AB} )+(1-\mu)P(a,0|A,z,\tau_{AB}'),
					 \end{eqnarray*}
					 and similarly,
					 \begin{eqnarray*}\label{element22}
						 &&{\nu}_{22}=\mu\; P(a,1|A,z,\tau_{AB} )+(1-\mu) P(a,1|A, z,\tau_{AB}').
					 \end{eqnarray*}
Note that, we have ${\nu}_{11}+{\nu}_{22}=\tr[\tilde{\rho}^{A}_a]=\mu P(a|A,\tau_{AB})+(1-\mu)P(a|A,\tau_{AB}').$
With the help of Eq. (\ref{ELHV}) and using
$\int P(a|A,\xi ) {P}_{\xi }d\xi=P(a|A,\tau_{AB}),$
$\int P'(a|A,\xi ) {P}_{\xi }d\xi=P(a|A,\tau_{AB}'),$
we have
$$
\left({\nu}_{11}+{\nu}_{22}\right)
=\int \wp(a|A,\xi)  P_{\xi }d\xi.
$$
 Because
 \begin{eqnarray*}
						 \tr[ \Pi^x_0\tilde{\rho}^{A}_a]=
						\frac{\nu_{11}+\nu_{12}}{2}+\textrm{Re}[{\nu}_{12}],
					 \end{eqnarray*}
					 with $\textrm{Re}[{\nu}_{12}]$ being the real part of ${\nu}_{12}$,
				 thus,

\begin{eqnarray*}
	\textrm{Re}[{\nu}_{12}]
&=&\int \left(\eta(x)-\frac{1}{2}\wp(a|A,\xi)\right)P_{\xi}d\xi.
\end{eqnarray*}

					 Similarly, because
					 \begin{eqnarray*}
						 \tr[{\Pi}^{y}_{0}\tilde{\rho}^{A}_a]
						 =\frac{\nu_{11}+\nu_{22}}{2}-\textrm{Im}[{\nu}_{12}],
					 \end{eqnarray*}
					 with $\textrm{Im}[{\nu}_{12}]$ being the imaginary part of
				 ${\nu}_{12}$, thus,
\begin{eqnarray*}
	-\textrm{Im}[{\nu}_{12}]
&=&\int \left(\eta(y)-\frac{1}{2}\wp(a|A,\xi)\right)P_{\xi}d\xi.
\end{eqnarray*}
On the other hand,
we have
\begin{eqnarray*}
	\frac{\nu_{11}-\nu_{22}}{2}
&=&\int \left(\eta(z)-\frac{1}{2}\wp(a|A,\xi)\right)P_{\xi}d\xi.
\end{eqnarray*}
Note that the following decomposition holds:
\begin{eqnarray*}\label{leftnew}
						 \tilde{\rho}^{A}_a&=&\frac{{\nu}_{11}+{\nu}_{22}}{2}\;\openone+ \textrm{Re}[{\nu}_{12}]
  \;\sigma_x\\
  &&-\textrm{Im}[{\nu}_{12}]
						 \;\sigma_y+\frac{{\nu}_{11}-{\nu}_{22}}{2}\;\sigma_z.\nonumber
					 \end{eqnarray*}

By combining the above equations, we finally deduce that Eq. (\ref{LHS3}) holds.
Thus, if there is a LHV model description for $\tau_{AB}$,
then there is a LHS model description for $\rho_{AB}$. This completes the proof.

 \textbf{ Remark 1. } Provided that the conditions in Theorem 1 are met,
  Theorem 1 actually provides a way to prove
  the following important property:
  if $\rho_{AB}$ is EPR steerable from A to B,
  then $\tau_{AB} $ is Bell nonlocal.
  Otherwise, if $\tau_{AB} $ is not Bell nonlocal,
there will be a LHS model for $\rho_{AB}$.

As a direct application of Theorem 1, we have Corollary 1.

\textbf{Corollary 1. }For any bipartite
qudit-qubit state $\tau_{AB}$ shared by Alice and Bob, define another state
\begin{eqnarray}\label{rhomu}
\rho_{AB}=\mu \; \tau_{AB} +(1-\mu) \tau'_{AB},
\end{eqnarray}
with $\tau'_{AB}={\tau }_{A}\otimes \frac{\openone+ c\sigma_3}{2}$, ${\tau }_{A}=\tr_{B}[\tau_{AB} ]$,
with the condition
$0\leq \mu \leq \frac{\sqrt{3}}{3}$ and
$
0\leq c \leq \frac{\sqrt{1-2\mu^2}-\mu}{1-\mu}.
$
If $\rho_{AB}$ is EPR steerable from A to B, then $\tau_{AB} $ is Bell nonlocal.

Proof. Assume that $\tau_{AB}$ is not Bell nonlocal, that is it has a LHV model.
Then we have
$$
        P(a, b|A, B, \tau_{AB})=\int P(a|A,\xi)P(b|B,\xi)P_{\xi}d\xi.
$$
Note that we have
        \begin{eqnarray*}
                P(a, b|A, B, \tau_{AB}')&=&\int P(a|A,\xi) \left( \frac{1}{2}(-1)^bcB_z+\frac{1}{2}  \right)  P_{\xi}d\xi.
        \end{eqnarray*}
        Let
        $$P'(a|A, \xi)=P(a|A,\xi), P'(b|B,\xi)=\left( \frac{1}{2}(-1)^bcB_z+\frac{1}{2}  \right)$$
        and substitute them into Eq.~(\ref{eqs:rxyz}) in Theorem 1,
we have ${r}_{x}=2\mu P(0|x,\xi)-\mu$, ${r}_y=2\mu P(0|y, \xi)-\mu$ and ${r}_z=2\mu P(0|z,\xi)+c-\mu c-\mu$.
To satisfy the assumption $|\vec{r}_{\xi }|\leq 1$ in Theorem 1,
it is equivalent to solving the following real quantifier elimination~\cite{Chen2} problem:
\begin{align*}
        &0\leq c\leq 1 \wedge 0\leq \mu\leq 1 \wedge (\forall P(0|x, \xi), P(0|y, \xi), P(0|z, \xi), \\
        &0\leq P(0|x, \xi)\leq 1 \wedge 0\leq P(0|y, \xi)\leq 1 \wedge 0\leq P(0|z,\xi)\leq 1\\
        &\implies {r}_x^2+{r}_y^2+{r}_z^2\leq 1).
\end{align*}
It is not hard to show that the solution is:
$0\leq \mu \leq \frac{\sqrt{3}}{3}$ and
$
0\leq c \leq \frac{\sqrt{1-2\mu^2}-\mu}{1-\mu}.
$
Since $\mu$ falls into the range required, the conditions of Theorem \ref{Theorem:Bell} are met.
Thus $\rho_{AB}$ has an LHS model, which is a contradiction.

\textbf{Remark 2. }
 This inspiring result clearly explore a curious quantum phenomenon: Bell nonlocal states can be constructed from steerable states. Such a novel finding
not only offers a distinctive way to study Bell's nonlocality without Bell's inequality but with steering
inequality, but also may avoid locality loophole in Bell's tests and make Bell's nonlocality easier for
demonstration. Interestingly, we can easily extract a simple corollary, namely Corollary 2, from Corollary 1,
which is a  well-known result derived in \cite{Chen,Cavalcanti2}.

\textbf{Corollary 2. }For any any bipartite
qudit-qubit state $\tau_{AB}$ shared by Alice and Bob, define another state
\begin{eqnarray}
\rho_{AB}=\mu \; \tau_{AB} +(1-\mu) \tau'_{AB},
\end{eqnarray}
with $\tau'_{AB}={\tau }_{A}\otimes \openone/2$, ${\tau }_{A}=\tr_{B}[\tau_{AB} ]=\tr_{B}[\rho_{AB} ]$
being the reduced density matrix at Alice's side,
and $\mu=\frac{1}{\sqrt{3}}$.
If $\rho_{AB}$ is EPR steerable from A to B, then $\tau_{AB} $ is Bell nonlocal. It is a special case of Corollary 1. When $d=2$, it will give us the derived results \cite{Chen,Cavalcanti2} for 2-qubit system.

 Proof. It is proved by setting $\mu=\frac{1}{\sqrt{3}}$ and $c=0$ in Corollary 1.

\begin{figure}
\begin{center}
\includegraphics[width=0.3\textwidth]{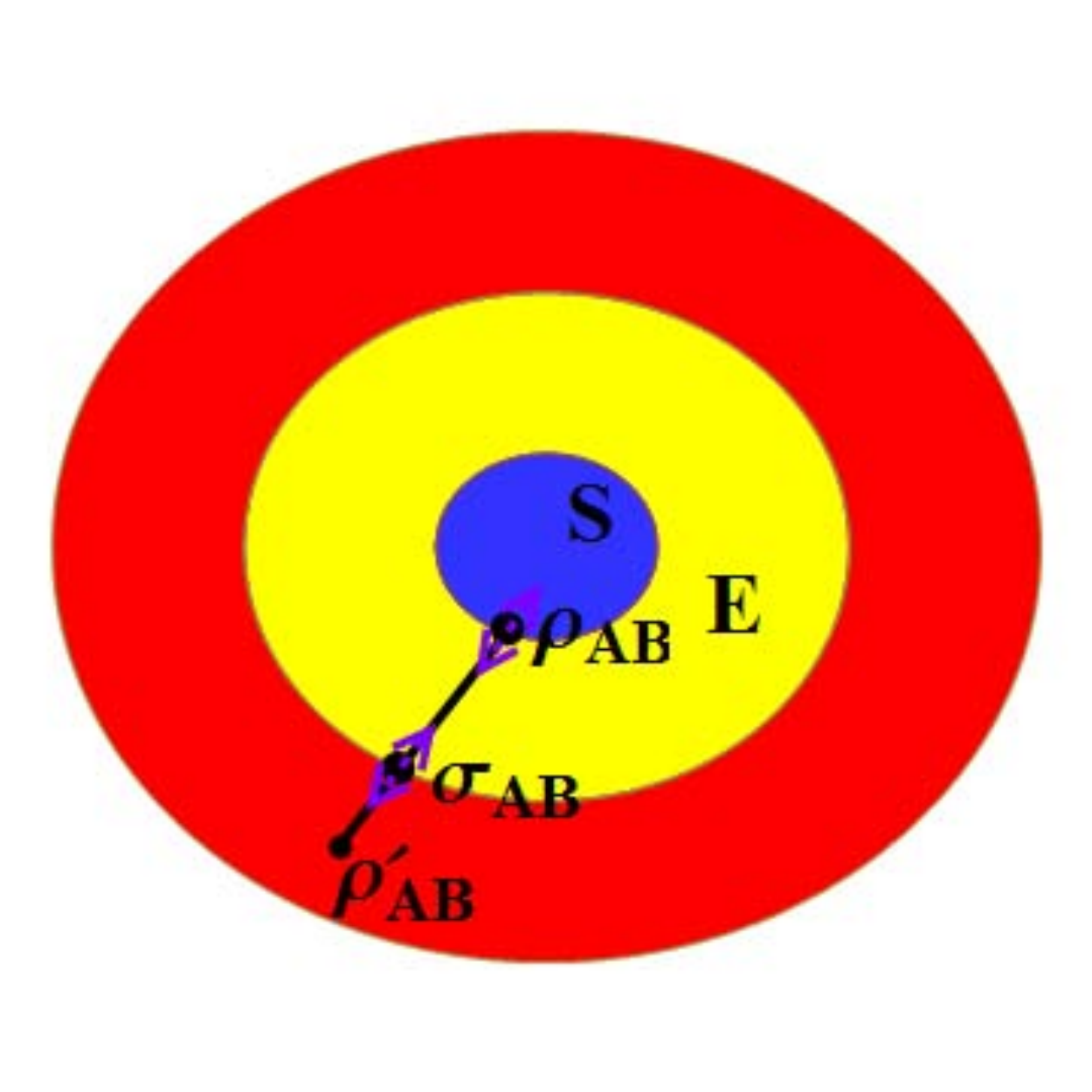}
 \caption{\label{fig2}(Color online) The Venn diagram of a mapping relation between quantum steering and quantum entanglement in Theorem 2. The states in Blue region is steerable. And those states in the Yellow region is unsteerable but entangled. All of the states which can be described by SPM are in the Red region. The state $\sigma_{AB}$ is mixed by an arbitrary separable state $\rho'_{AB}$ and the other arbitrary state $\rho_{AB}$. Theorem 2 gives a mapping criterion between $\rho_{AB}$ and $\sigma_{AB}$. The purple arrows mean that, if $\sigma_{AB}$ is entangled, then $\rho_{AB} $ is EPR steerable, equivalently if $\rho_{AB} $ is unsteerable then $\sigma_{AB}$ is separable.}
\end{center}
\end{figure}

\emph{Mapping criterion between quantum steering and quantum entanglement.-} Similarly, a mapping criterion between quantum steering and quantum entanglement can be precisely derived. It is shown that steerable states can be constructed from some entangled states, which indicates that EPR steering can be detected indirectly through entanglement (see Fig. (2)), and offers a distinctive way to study EPR steering. The result can be expressed as the following
theorem.
 \textbf{Theorem 2. } In a bipartite qubit-qudit system, we define a map ${\cal N}: \rho_{AB}\rightarrow \mu \; \rho_{AB} +(1-\mu) \rho'_{AB}$, $0\leq \mu\leq 1,$ where
  $\rho_{AB}$ is an arbitrary bipartite qubit-qudit state  shared by Alice and Bob,
  while $\rho_{AB}'$ is a bipartite qubit-qudit state constructed in such a way that whenever $\rho_{AB}$ has a LHS model:
  $$
        P(a, b|A, B, \rho_{AB})=\int P(a|A,\xi)P_Q(b|B,\xi)P_{\xi} d\xi,
  $$
  $\rho_{AB}'$ also has a LHS model:
  $$
  P(a, b|A, B, \rho_{AB}')=\int P'(a|A,\xi)P_Q(b|B,\xi)P_{\xi} d\xi.
  $$
  Note that the above two equations have the same $P_Q(b|B,\xi)$ and $P_{\xi}$.
  If there exists a range of $\mu$ such that $r_x^2+r_y^2+r_z^2\leq 1$ holds
        for any probability distributions $0\leq P(0|A,\xi)\leq 1$, where $A\in\{x, y, z\}$, and
	\begin{eqnarray}
          \label{eqn:lhsm-sm}
		r_x&=2(\mu P(0|x,\xi) + (1-\mu)P'(0|x,\xi))-1,\nonumber\\
		r_y&=2(\mu P(0|y,\xi) + (1-\mu)P'(0|y,\xi))-1,\nonumber\\
		r_z&=2(\mu P(0|z,\xi) + (1-\mu)P'(0|z,\xi))-1,
	\end{eqnarray}
  then when $\mu$ falls into this range,
        one can construct a SPM for $\sigma_{AB}={\cal N}(\rho_{AB})$.


Proof.	To prove that $\sigma_{AB}$ has a SPM description is equivalent to proving the following equation has a solution:
	\begin{equation}
	\label{eq:sm}
	P(a,b|A,B,\sigma_{AB})=\int \wp_Q(a|A,\xi)\wp_Q(b|B,\xi)\wp_{\xi}d\xi.
	\end{equation}
        A solution is given by
        $$
        \wp_Q(a|A,\xi)=\tr[\Pi_a^A\rho_{\xi}^A], \wp_Q(b|B,\xi)=P_Q(b|B,\xi), \wp_{\xi}=P_{\xi},
        $$
        where $\rho_{\xi}^A=\frac{\openone+\vec{\sigma}\cdot\vec{r}_{\xi}^A}{2}$,
        and $\vec{r}_{\xi}^A=(r_x,r_y,r_z)$ with $r_x,r_y,r_z$ given in Eq.~(\ref{eqn:lhsm-sm}).
        The assumption that $|\vec{r}_{\xi }|\leq 1$ ensures that $\rho_\xi$ is a density matrix.

Next we prove that the above solution makes Eq.~(\ref{eq:sm})~hold.
It is easy (by hand or a computer algebra system) to check that
\begin{align*}
	\wp_Q(a|A,\xi)&=\frac{1+(-1)^aA\cdot\vec{r}_{\xi}^A}{2}.
\end{align*}
Thus we have
\begin{equation}
  \label{eq:lhs-sp-rh}
\begin{array}{rl}
   &\int \wp_Q(a|A,\xi)\wp_Q(b|B,\xi)\wp_{\xi}d\xi\\
  =&\frac{1}{2}\int P_Q(b|B,\xi)P_{\xi}d\xi\\
  & + \frac{(-1)^a}{2}\int (A_xr_x+A_yr_y+A_zr_z)P_Q(b|B,\xi)P_{\xi} d\xi
\end{array}
\end{equation}
On the other hand,
we have
$$
\begin{array}{rl}
   &\int A_xr_xP_Q(b|B,\xi)P_{\xi}d\xi\\
  =& 2\mu A_xP(0, b|x, B,\rho_{AB})+2(1-\mu)A_xP(0, b|x, B,\rho'_{AB})\\
  &-A_x\int P_Q(b|B,\xi)P_{\xi} d\xi, \\
\end{array}
$$
and
$$
\begin{array}{rl}
  &\frac{(-1)^a}{2} A_xP(0, b|x, B,\rho_{AB})\\
 =& \frac{1}{2}\tr[(\frac{(-1)^aA_x\openone+(-1)^aA_x\sigma_x}{2})\otimes\Pi_b^{B})\rho_{AB} ],
\end{array}
$$
and
$$
\tr[(\openone\times\Pi_b^B)\rho_{AB}]=\int  P_Q(b|B,\xi)P_{\xi} d\xi,
$$
and
$$
\tr[(\frac{\openone+(-1)^aA}{2}\otimes \Pi_b^{B})\rho_{AB}]=P(a,b|A,B,\rho_{AB}).
$$
Thus the following holds:
\begin{equation}
  \label{eq:lhs-sp-key}
\begin{array}{rl}
  &\frac{(-1)^a}{2}\int (A_xr_x+A_yr_y+A_zr_z)P_Q(b|B,\xi)P_{\xi} d\xi\\
  =&\mu P(a,b|A,B,\rho_{AB})+(1-\mu)P(a,b|A,B,\rho'_{AB})\\
   &-\frac{1}{2}\int P_Q(b|B,\xi)P_{\xi} d\xi.
\end{array}
\end{equation}
Since
\begin{align*}	
	P(a, b|A, B, \sigma_{AB})
				 &=\mu P(a, b|A, B, \rho_{AB})\\&+(1-\mu)P(a, b|A, B, \rho_{AB}'),
\end{align*}
combining with Eq. (\ref{eq:lhs-sp-rh}) and Eq. (\ref{eq:lhs-sp-key}),
we have Eq. (\ref{eq:sm}).
This proves the theorem.


 \textbf{Remark 3. } Provided the conditions in Theorem 2 are met,
  Theorem 2 provides a way to prove the following important property:
  if $\sigma_{AB}$ is entangled,
  $\rho_{AB} $ is EPR steerable
  in the sense that Alice can steer Bob.
  Otherwise, if $\rho_{AB} $ is not EPR steerable from A to B,
there will be a SPM description for $\sigma_{AB}$.

As a direct application of Theorem 2,
we have the following result.

\textbf{Corollary 3. }For an arbitrary bipartite qubit-qudit state $\rho_{AB}$ shared by Alice and Bob, define

\begin{eqnarray}\label{rhomu}
\sigma_{AB}=\mu \; \rho_{AB} +(1-\mu) \rho'_{AB},
\end{eqnarray}
with $\rho'_{AB}=\frac{\openone + c\sigma_3}{2}\otimes \rho_B$, ${\rho }_{B}=\tr_{A}[\rho_{AB} ]$,
with the condition
$0\leq \mu \leq \frac{\sqrt{3}}{3}$ and
$
0\leq c \leq \frac{\sqrt{1-2\mu^2}-\mu}{1-\mu}.
$
If $\sigma_{AB}$ is entangled state,
then $\rho_{AB}$ is the steerable state in the sense that Alice can steer Bob.

 Proof. Assume that $\rho_{AB}$ is not steerable, that is it has a LHS model:
  $$
  P(a, b|A, B, \rho_{AB})=\int P(a|A,\xi)P_Q(b|B,\xi)P_{\xi} d\xi.
  $$
  Since the marginal probability satisfies
  $$
  P(b|B, \rho_B)=\int P_Q(b|B,\xi)P_{\xi}d\xi,
  $$
  we have
  $$
  P(a, b|A, B, \rho_{AB}')=\int \left( \frac{1+(-1)^acA_z}{2}  \right)P_Q(b|B,\xi)P_{\xi}d\xi.
  $$
   Set $P'(a|A,\xi)=\frac{1+(-1)^acA_z}{2}$ and substitute it into Eq. (\ref{eqn:lhsm-sm}),
   we have
   \begin{align*}
     {r}_{x}&=2\mu P(0|x,\xi)-\mu,\\
     {r}_y&=2\mu P(0|y, \xi)-\mu,\\
     {r}_z&=2\mu P(0|z,\xi)+c-\mu c-\mu.
   \end{align*}

   To satisfy the assumption $r_x^2+r_y^2+r_z^2\leq 1$ in Theorem 2,
   it is equivalent to solving exactly the same real quantifier elimination problem as the one in
   Corollary 1,
   whose solution is
   $0\leq \mu \leq \frac{\sqrt{3}}{3}$ and
$
0\leq c \leq \frac{\sqrt{1-2\mu^2}-\mu}{1-\mu}.
$
Since $\mu$ falls into the range required, the conditions of Theorem 2  are met.
Thus $\sigma_{AB}$ has an SPM, which is a contradiction.

\textbf{Corollary 4. }For an arbitrary bipartite qubit-qudit state $\rho_{AB}$ shared by Alice and Bob, one can map it into
	a new state defined by: $\sigma_{AB}=\mu\rho_{AB}+(1-\mu)\rho_{AB}'$,
	with $\rho_{AB}'=\frac{\openone}{2}\otimes \rho_{B}$,
	where $\rho_{B}={\tr}_A[\rho_{AB}]$, $\mu=1/\sqrt{3}$,
	if $\sigma_{AB}$ is entangled state, then $\rho_{AB}$
	is the steerable state in the sense that Alice can steer Bob. When $d=2$, it will reduce to 2-qubit system.

 Proof. It can be deduced directly from Corollary 3 by setting $\mu=\frac{1}{\sqrt{3}}$ and $c=0$.


\emph{Conclusion.-} We not only presented a mapping criterion between Bell nonlocality and quantum steering, but also a mapping criterion between quantum steering and quantum entanglement, starting from these fundamental concepts of quantum correlations.
Many novel quantitative results on the relation of such quantum correlations were derived.
It is shown that part of these previous known result in \cite{Chen,Cavalcanti2}
is only a special case in our simple mapping criterion.
Our result not only pinpoints a
deep connection among quantum entanglement, quantum steering and Bell nonlocality,
but also provides a feasible approach to experimentally test a difficultly-verified quantum correlation
by translating it into an easily-verified problem.

The method we use in the present paper provides a particularly new perspective to understand various quantum correlations,
and shines light on the intricate relations among them.
As we showed with concrete examples, this connection allows us to translate results from one concept to another.
There is no doubt that this method is easily-extendable, so, for future work it would be very interesting to
use such method to explore many different mapping criteria especially in higher dimensions. Another open question is that, such kind of mapping criterion between Bell nonlocality and quantum entanglement is still unknown. If such mapping criterion exists, which indicates that Bell nonlocality can be detected indirectly through quantum entanglement, definitely, it will supply a distinctive way to avoid locality loophole in Bell tests and make Bell nonlocality easier for demonstration. Hence, this open question is also important enough to deeply explore.

\emph{Acknowledgement.-}
The authors would like to thank anonymous referees for valuable comments and suggestions.
C.B.C. is supported by the National Natural Science Foundation of China (No. 11771421, 11471307, 61572024, 11671377),
the Key Research Program of Frontier Sciences of CAS (QYZDB-SSW-SYS026) and EU H2020-FETOPEN-2016-2017-CSA project $SC^{2}$ (712689).
C.L.R. was supported by National key research and development program (No. 2017YFA0305200), the Youth Innovation Promotion Association (CAS) (No. 2015317), the National Natural Science Foundation of China (No. 11605205), the Natural Science Foundation of Chongqing (No. cstc2015jcyjA00021,cstc2018jcyjA2509), the Entrepreneurship and Innovation Support Program for Chongqing Overseas Returnees (No. cx2017134, cx2018040), the fund of CAS Key Laboratory of Quantum Information. J.L.C. is supported by National Natural Science Foundations of China (Grant Nos. 11475089 and 11875167).

C. B. Chen and C. L. Ren contributed equally to this work.

\end{document}